\documentclass[aps,prl,twocolumn,superscriptaddress,showpacs]{revtex4} 
\usepackage{graphicx}
\usepackage{amsmath}
\begin{document}

\newcommand{\<}{\langle}
\renewcommand{\>}{\rangle}
\newcommand{\Lmin}{L_{\rm min}}

\title{Critical Scaling Properties at the 
Superfluid Transition of $^4$He in Aerogel}

\author{Marios Nikolaou} \affiliation{Department of Theoretical
Physics, Royal Institute of Technology, AlbaNova, SE-106 91 Stockholm,
Sweden}

\author{Mats Wallin} \affiliation{Department of Theoretical Physics,
Royal Institute of Technology, AlbaNova, SE-106 91 Stockholm, Sweden}

\author{Hans Weber} \affiliation{Department of Physics, Lule{\aa}
University of Technology, SE-971 87 Lule{\aa}, Sweden}

\date{\today}

\begin{abstract}
We study the superfluid transition of $^4$He in aerogel by Monte Carlo
simulations and finite size scaling analysis.  Aerogel is a highly
porous silica glass, which we model by a diffusion limited cluster
aggregation model.  The superfluid is modeled by a three dimensional
XY model, with excluded bonds to sites on the aerogel cluster.  We
obtain the correlation length exponent $\nu=0.73 \pm 0.02$, in
reasonable agreement with experiments and with previous simulations.
For the heat capacity exponent $\alpha$, both experiments and previous
simulations suggest deviations from the Josephson hyperscaling
relation $\alpha=2-d\nu$.  In contrast, our Monte Carlo results
support hyperscaling with $\alpha= -0.2\pm 0.05$.  We suggest a
reinterpretation of the experiments, which avoids scaling violations
and is consistent with our simulation results.
\end{abstract}


\maketitle

Experiments on $^4$He immersed in ultralight aerogel, forming highly
porous fractal structures, have demonstrated new universality classes
of the superfluid phase transition \cite{chan88,yoon}.  Striking
deviations of the critical exponents from the bulk lambda transition
appear to arise due to fractal correlations in the aerogel, but the
details are not well understood.  Furthermore, the new critical
exponents tend to violate common scaling laws, such as the Josephson
hyperscaling relation.  A theoretical study of the unusual scaling
properties of the superfluid transition in fractal random media is
motivated in order to better understand these surprising results.

Experiments on helium in aerogel use ultralight aerogel samples with a
highly porous structure that consists of silica strands of about 5 nm
in diameter and a distance between neighboring strands of up to 200 nm
\cite{yoon}.  The porosity can be varied in the fabrication
process. Small-angle x-ray scattering indicates a fractal structure
extending from a couple of nanometers up to more than 100 nm
\cite{chan96}.  The superfluid density of the $^4$He has been measured
by a torsional oscillator technique, and the heat capacity by an ac
heating method \cite{yoon}.  The superfluid density was found to
vanish at the transition as a power law $\rho_s=\rho_0 |t|^\zeta$,
where $t=1-T/T_c$ and $\zeta\approx 0.79$ for $m=5\%$ volume fraction
of aerogel, $\zeta \approx 0.76$ for $m=2\%$, and $\zeta\approx 0.72$
for $m=0.5\%$.  These values are significantly greater than
$\zeta\approx 0.67$ found in bulk $^4$He.  The heat capacity has a
peak at $T=T_c$ that was fitted to the form $C =
A_{\pm}|t|^{-\alpha_\pm}+B$, giving $\alpha_+ \ne \alpha_-$.
Alternatively, assuming that $\alpha_+=\alpha_-=\alpha$ gives $\alpha
\approx -0.57$ for $m=5\%$ aerogel, $\alpha \approx -0.39$ for
$m=2\%$, and $\alpha \approx -0.13$ for $m=0.5\%$ \cite{yoon}.
Surprisingly, these exponents depend on the aerogel porosity, and
deviate considerably from the value obtained from the Josephson
hyperscaling relation, $\alpha=2-d\nu$, with $d=3$ and $\nu=\zeta$.

The $^4$He transition in aerogel has been studied by computer
simulation in Refs.\ \cite{moon,vasquez}, and we follow many aspects
of these works.  Moon and Girvin \cite{moon} studied a three
dimensional (3D) XY model with aerogel modeled as a connected
percolation fractal cluster with fractal dimension $D_f\approx 2.5$.
They obtained the expected scaling behavior of the superfluid density,
i.e., $\rho_s \sim \xi^{2-d}$ with $d=3$, and a correlation length
that diverges as $\xi \sim |T-T_c|^{-\nu}$ with exponent $\nu\approx
0.722$, in agreement with experiments \cite{yoon}.  However, they did
not study the unusual heat capacity exponents.  V\'asquez et al.\
\cite{vasquez} also studied a 3D XY model, but with aerogel modeled
with a diffusion limited cluster aggregation (DLCA) algorithm,
generating a fractal cluster with dimension $D_f\approx 1.8$
\cite{meakin84}.  They find porosity dependent exponents similar to
those from the experiments.  Specifically, the heat capacity exponent
displays a similar violation of hyperscaling as obtained from
experiments.

In this Letter we study the superfluid transition in aerogel by Monte
Carlo (MC) simulations and finite size scaling analysis, using a more
extensive disorder averaging and larger system sizes than in previous
simulations.  In contrast to results from experiments and earlier
simulations, we obtain critical exponents that are both porosity
independent and obey usual scaling laws.  In particular, our MC data
is consistent with the Josephson hyperscaling relation.  We reanalyze
the experimental data for the heat capacity from Ref.\ \cite{yoon} in
light of our new scaling results, and argue that the formula $C =
A_{\pm}|t|^{-\alpha_\pm}+B$ used in the previous data analysis may not
directly apply for this problem.  We present an alternative analysis
that leads to consistency between experiments and our simulation
results.

{\em DLCA model. --} Following Ref. \cite{vasquez}, we model the
aerogel using a DLCA algorithm \cite{meakin,kolb}.  We consider a
simple cubic lattice with $L \times L \times L$ sites and periodic
boundary conditions in all directions.  As a first step $mL^3$
``particles'' are distributed on random lattice sites, where $m$ is
the volume fraction of the aerogel.  Particles on nearest neighbor
lattice sites form clusters, whose structure is kept fixed.  Clusters
are allowed to diffuse by rigid translations, using an iterative DLCA
algorithm \cite{hasmy}.  If two clusters meet they connect
irreversibly and form a larger cluster.  The iteration continues until
a single cluster remains, which is used to model the aerogel, and
plays the role of correlated disorder for the helium.  The aerogel
volume fractions considered here are $m=5\%$ and $m=10\%$.

{\em 3D XY model. --} The universal properties of the superfluid
transition of $^4$He are captured by the 3D XY model
given by
\begin{equation}
H=-\sum_{\<i,j\>} J_{ij} \cos(\theta_i-\theta_j),
\label{H}
\end{equation}
where $\<i,j\>$ denotes nearest neighbor sites.  The coupling constant
$J_{ij}$ is set to $J=1$ between pairs of $^4$He sites, and $J=0$ if
one or both lattice sites belong to the DLCA cluster.  The excluded
bonds act like quenched disorder for the 3D XY model, and $\theta_i$ is
the phase of the superfluid order parameter at site $i$ occupied by
helium.

{\em Monte Carlo simulations. --} Our MC simulations use the
collective Wolff update method \cite{wolff}, combined with the
temperature exchange method \cite{exchange}, which reduces critical
slowing down of the simulation.  An exchange update attempt was
performed after every ten Wolff cluster updates.  For each random bond
realization we discard between $2\times 10^4$ and $2\times 10^5$ Wolff
clusters to approach equilibrium, depending on system size, followed
by the same number of updates collecting data.  In the temperature
exchange method we simulate a set of different temperatures in
parallel, and allow MC moves exchanging the temperatures between the
different configurations.  The results were averaged over $5\times
10^3$ DLCA clusters for $L \le 50$, and for $2.5\times 10^3$ clusters
for $L \ge 60$.  We carefully checked for convergence of the
simulation by increasing the number of initial discarded update steps
until stable results were obtained.

{\em Calculated quantities and finite size scaling relations. --} The
superfluid density $\rho_s$ is proportional to the helicity modulus
$Y$, which gives the increase in free energy density as $\Delta
f=\frac{1}{2} Y \Delta^2$ in the presence of a uniform phase twist
$\Delta$ imposed across the system, here taken in the $x$ direction
\cite{Fisher_PRA73}.  The helicity modulus is then given by
\cite{teitel90}
\begin{eqnarray}
Y(T,L)=\frac{1}{L^3} [ \< -\sum_j J_j^x \cos(\theta_{j+\hat{e}_x}-\theta_j)
  \> ] 
\nonumber
\\
- \frac{1}{L^3T} [ \< ( \sum_j J_j^x
  \sin(\theta_{j+\hat{e}_x}-\theta_j) )^2 \> ],
\label{Y}
\end{eqnarray}
where $J_j^x=J_{j+\hat{e}_x,j}$.  Here $\< \cdots \>$ denotes thermal
average, and $[ \cdots ]$ denotes average taken over different
realizations of DLCA clusters.

In order to analyze MC data for finite systems of size $L^3$, we start
from the finite size scaling relation for the singular part of the
free energy density, $f_s/T=L^{-d} \tilde{f}_\pm (L/\xi)$, where
$d=3$, $\tilde{f}_\pm$ are scaling functions, and $\xi\sim
|T-T_c|^{-\nu}$ is the correlation length \cite{goldenfeld}.  The
corresponding scaling result for a phase gradient is $\Delta = L^{-1}
\tilde{\Delta}_\pm (L/\xi)$.  The helicity modulus becomes $Y \sim
L^{2-d}$, or, alternatively, $Y \sim |T-T_c|^\zeta$ \cite{Fisher_PRA73}
with $\zeta=(d-2)\nu$.  For $d=3$ the finite size scaling relation
becomes \cite{moon}
\begin{equation}
LY/T=(1+aL^{-\omega}) \tilde{Y}(L^{1/\nu}t),
\label{Yscaling}
\end{equation}
where $\tilde{Y}$ is a scaling function to be determined below from MC
data, and $t=T-T_c$.  The term $aL^{-\omega}$ is an amplitude
correction to scaling that will be discussed below \cite{campostrini}.

The Josephson hyperscaling relation for the heat capacity exponent,
$\alpha=2-d\nu$, follows from $c\sim \partial^2 f_s/\partial T^2 \sim
\partial^2 t^{d\nu}/\partial t^2 \sim t^{d\nu-2} = t^{-\alpha}$
\cite{goldenfeld}.  The heat capacity is given by $c= \left[ \<H^2\> -
\<H\>^2 \right]/(L^3T^2)$, and the finite size scaling relation is
\cite{schultka}
\begin{equation}
c=L^{\alpha/\nu} \tilde{c}(L^{1/\nu}t)+b,
\label{c}
\end{equation}
where $b$ represents the nonsingular, analytic contribution to the
heat capacity, which depends on temperature but not system size.
Since we focus on a narrow temperature interval around the transition,
we take $b$ to be constant.  Alternatively, the scaling function can
be determined directly by eliminating $b$:
\begin{equation}
\tilde{c}(x)=\frac{c(x,L_1)-c(x,L_2)}{L_1^{\alpha/\nu}
-L_2^{\alpha/\nu}},
\label{ctilde}
\end{equation}
where $x$ is the scaling variable $L^{1/\nu}t$ and $L_1,L_2$ denote two
different system sizes.  These two formulas gave the same results for
the critical exponents.

\begin{figure}
\includegraphics[width=0.9\columnwidth]{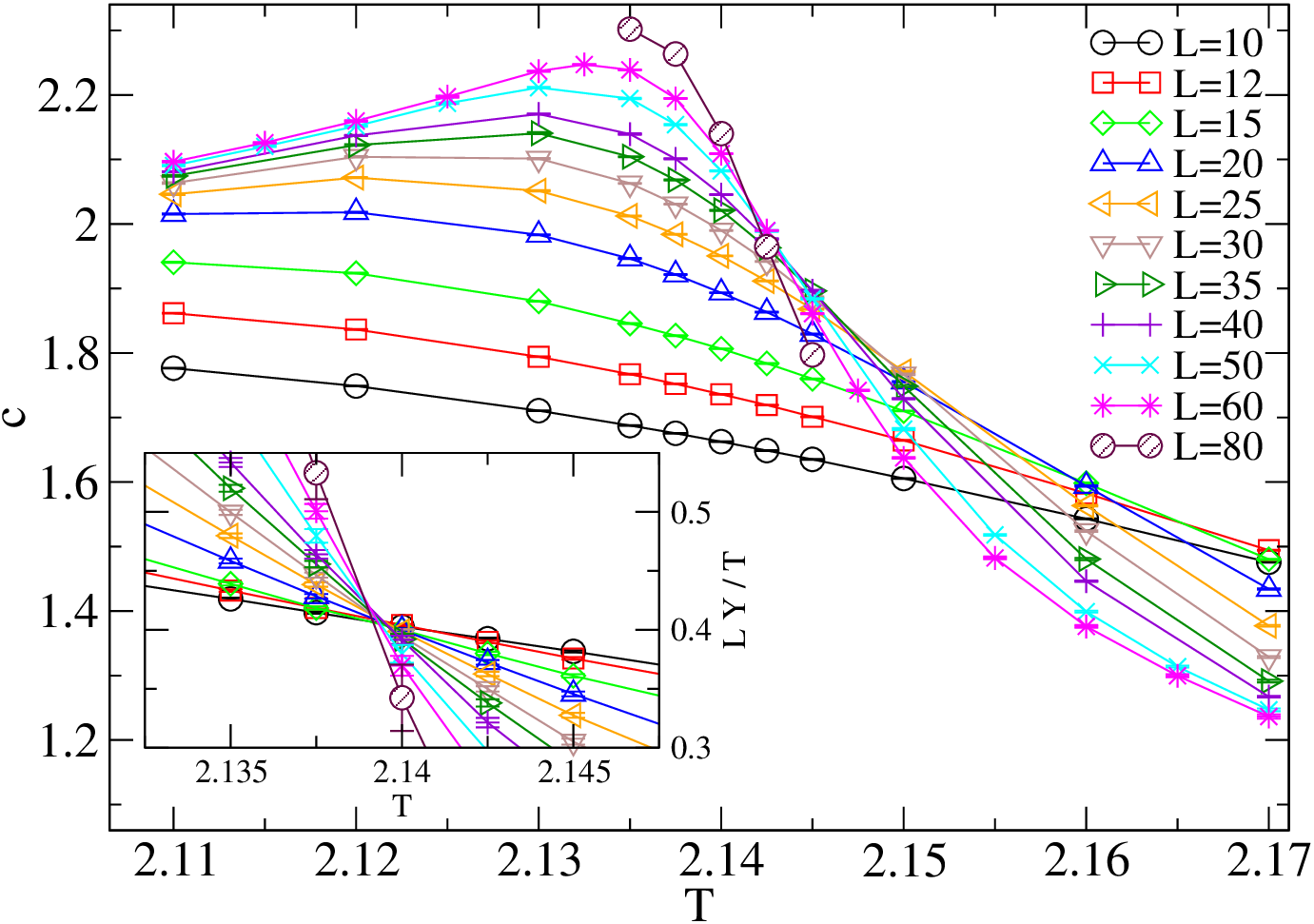}
\caption[]{(color online) MC data for the heat capacity $c$ (main
figure) and the dimensionless helicity modulus $LY/T$ (inset) as a
function of temperature $T$ for different system sizes $L$, with
$m=5\%$ aerogel volume fraction.  }
\label{fig:Yc}
\end{figure}

{\em Results. --} Figure \ref{fig:Yc} shows MC data for the heat
capacity $c$ (main figure) and the dimensionless helicity modulus
$LY/T$ (inset) vs temperature $T$ for a set of different system sizes
$L$ for an aerogel volume fraction of $m=5\%$.  According to the
scaling relation Eq.\ (\ref{Yscaling}), the dominant L dependence for
the helicity modulus at $T=T_c$ is $Y \sim L^{-1}$, showing that the
superfluid transition temperature $T_c$ is where all data curves for
$LY/T$ for different system sizes $L$ intersect.  However, a close
examination of the data curves for $Y$ in Fig.\ \ref{fig:Yc} reveals a
small systematic drift in the intersection temperatures between
successive pairs of system sizes, which complicates an accurate
determination of $T_c$.  The drift can be included in the scaling
analysis by assuming an amplitude correction to scaling of the form
included above in Eq.\ (\ref{Yscaling}).  We found that varying
$\omega$ in the interval from 0.6 to 1 leads to practically
indistinguishable fits to the MC data, with nearly the same values for
the other exponents.  The precision of our MC data is not enough for a
more accurate estimate of $\omega$, and the results shown below are
for the choice $\omega=1$.  For the heat capacity no correction to
scaling was found to be necessary.

The following method produces a finite size data collapse of MC data
for the helicity modulus by fitting to Eq.\ (\ref{Yscaling}).  The
$\chi^2$ value of the fit is obtained as $\chi^2 = (1/n) \sum_{i>j}
\int dx [\tilde{Y}_i(x)-\tilde{Y}_j(x)]^2$, where $i,j$ refer to $n$
pairs of system sizes $L_i,L_j$.  The integrals are evaluated using
cubic spline interpolation at 20 evenly spaced points within the $x$
interval in which the pair of functions in the integrand overlap.  The
fit parameters are determined by minimizing $\chi^2$ over a fine
multidimensional grid of values, whose resolution is finer than the
error estimates on the parameters.  Data points at small values of $Y$
with poor statistics are excluded from the fits.  A similar procedure
is used to fit Eq.\ (\ref{c}) to the heat capacity data.  We tried
both independent scaling fits for $Y$ and $c$, and joint fits to scale
both quantities simultaneously, which we found to be more stable.

\begin{figure}
\includegraphics[width=\columnwidth]{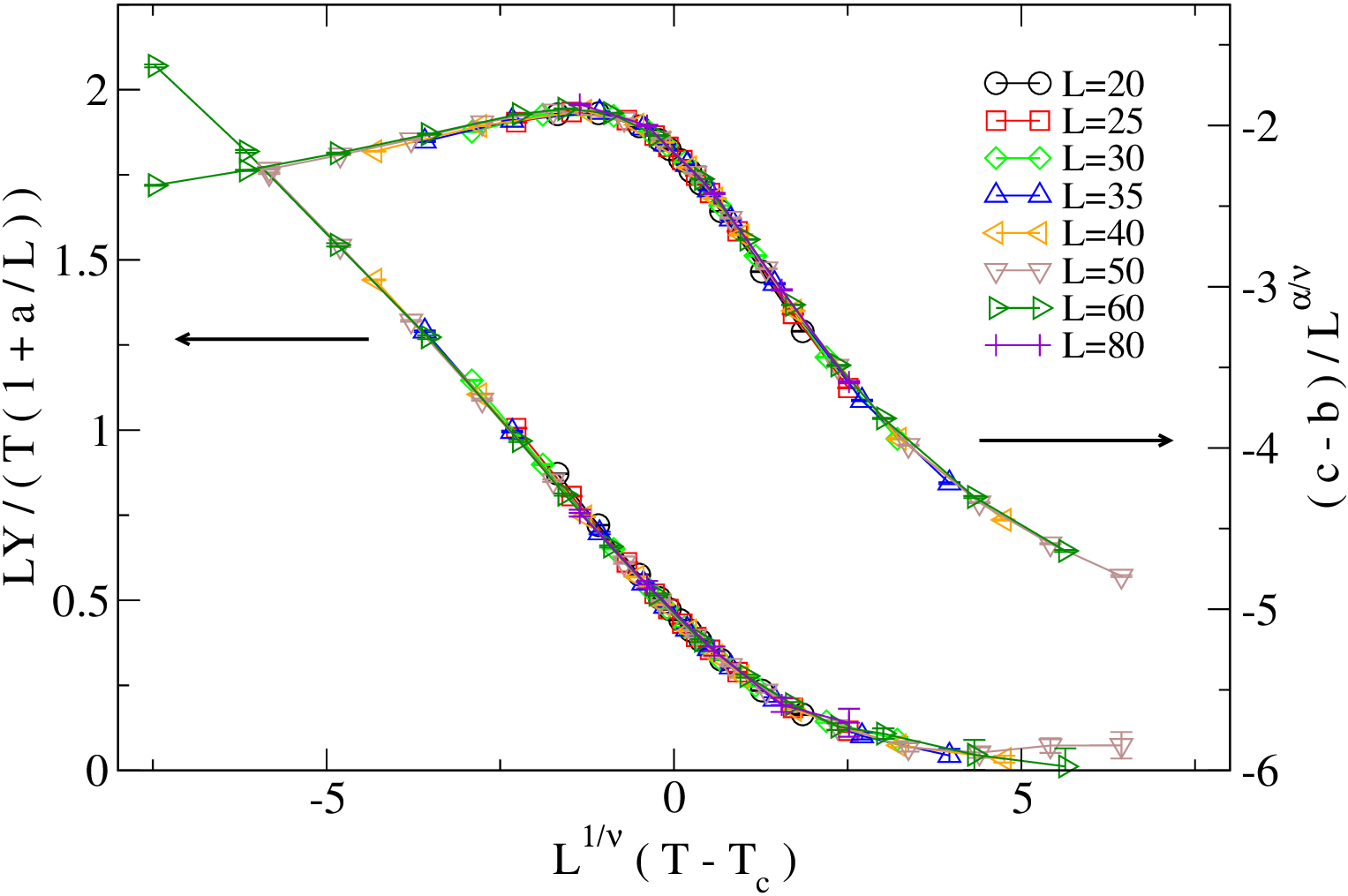}
\caption[]{(color online) Finite size scaling data collapses of MC
data showing the scaling functions for the helicity modulus $Y$ and
the heat capacity $c$ for $m=5\%$.  The data collapses giving the
scaling functions are obtained from a joint fit to Eqs.\
(\ref{Yscaling}) and (\ref{c}) for system sizes $L\ge 20$, assuming
that $\alpha$ is given by the hyperscaling relation $\alpha=2-d\nu$.
}
\label{fig:Ycscaling}
\end{figure}

To determine the critical exponents $\nu$ and $\alpha$ we attempt some
different approaches.  We first assume that the hyperscaling relation
$\alpha=2-d\nu$ is valid, which leaves the fit parameters
$T_c,\nu,a,b$.  The result of joint fits of Eqs.\ (\ref{Yscaling}) and
(\ref{c}) to MC data is shown in Fig.\ \ref{fig:Ycscaling} for
$m=5\%$.  As seen in the figure the fit produces good data collapses
both for $Y$ and for $c$.  The fit parameters for $m=5$\% are:
$T_c=2.1385 \pm 0.0005, \nu=0.73 \pm 0.02, a=-1.9, b=2.85$, and for
$m=10$\% (data not shown): $T_c=2.0450 \pm 0.0005, \nu=0.73 \pm 0.02,
a=-2.0, b=2.6$.  Error bars are estimated from the variation in the
parameter values upon varying the range of system sizes included in
the fits.  We observe a slight tendency for the exponents to drift
towards the 3D XY results upon increasing the system sizes included in
the scaling fits, but larger system sizes would be required to study
this effect further.  The values of $T_c$ are close to the value
$T_c\approx 2.203$ for the pure 3D XY model, which is expected since
the DLCA cluster fills only a tiny fraction of the entire system
volume.  The values of $\nu$ for 5\% and 10\% aerogel closely agree
with each other, and we get the corresponding heat capacity exponent
$\alpha=2-3\nu= -0.20 \pm 0.05$, independent of the aerogel porosity.
We also tried joint scaling fits without assuming the hyperscaling
relation for $\alpha$, but instead treating $\nu$ and $\alpha$ as
independent fit parameters, with similar results.  On the other hand,
using the parameter values suggested by the experiments \cite{yoon}
gives a good fit to the helicity modulus, but a very poor fit to the
heat capacity data.  Thus, from fits to our MC data we conclude that
the exponents are porosity independent and obey hyperscaling, in
contrast to suggestions in Refs.\ \cite{vasquez,yoon}.

{\em Comparison with experiments. --} Here we compare our critical
exponents to the experimental results in Ref.\ \cite{yoon}.  Our
scaling of the helicity modulus gives the correlation length exponent
$\nu \approx 0.73$, which compares quite well with the experimental
values $\nu=0.72-0.79$ from torsional oscillator measurements of the
superfluid density.  However our heat capacity exponent $\alpha
\approx -0.20$ is far from the experimental values $\alpha =-0.39$ for
$m=2\%$ aerogel, and $\alpha=-0.57$ for $m=5\%$.  These results are
clearly incompatible.

We now argue that our scaling results can actually describe the
universal scaling function probed by the experiment.  However, we also
suggest that the experimental data contains nonuniversal features
that must be included in the data analyses as well.  We first assume
the presence of a finite cutoff length scale in the experimental
system.  Figure 3 of Ref.\ \cite{yoon} shows that the data for the
$^4$He transition without any aerogel produces a sharp lambda curve.
However, the presence of the aerogel seems to slightly round off the
maximum of the heat capacity curves, compared to the sharp peak
predicted by the formula $c \sim |T-T_c|^{-\alpha}$.  Rounding of the
experimental curves therefore seems to be caused by the presence of
the aerogel, and suggests a finite length scale that may be associated
with the typical aerogel pore size.

\begin{figure}
\includegraphics[width=\columnwidth]{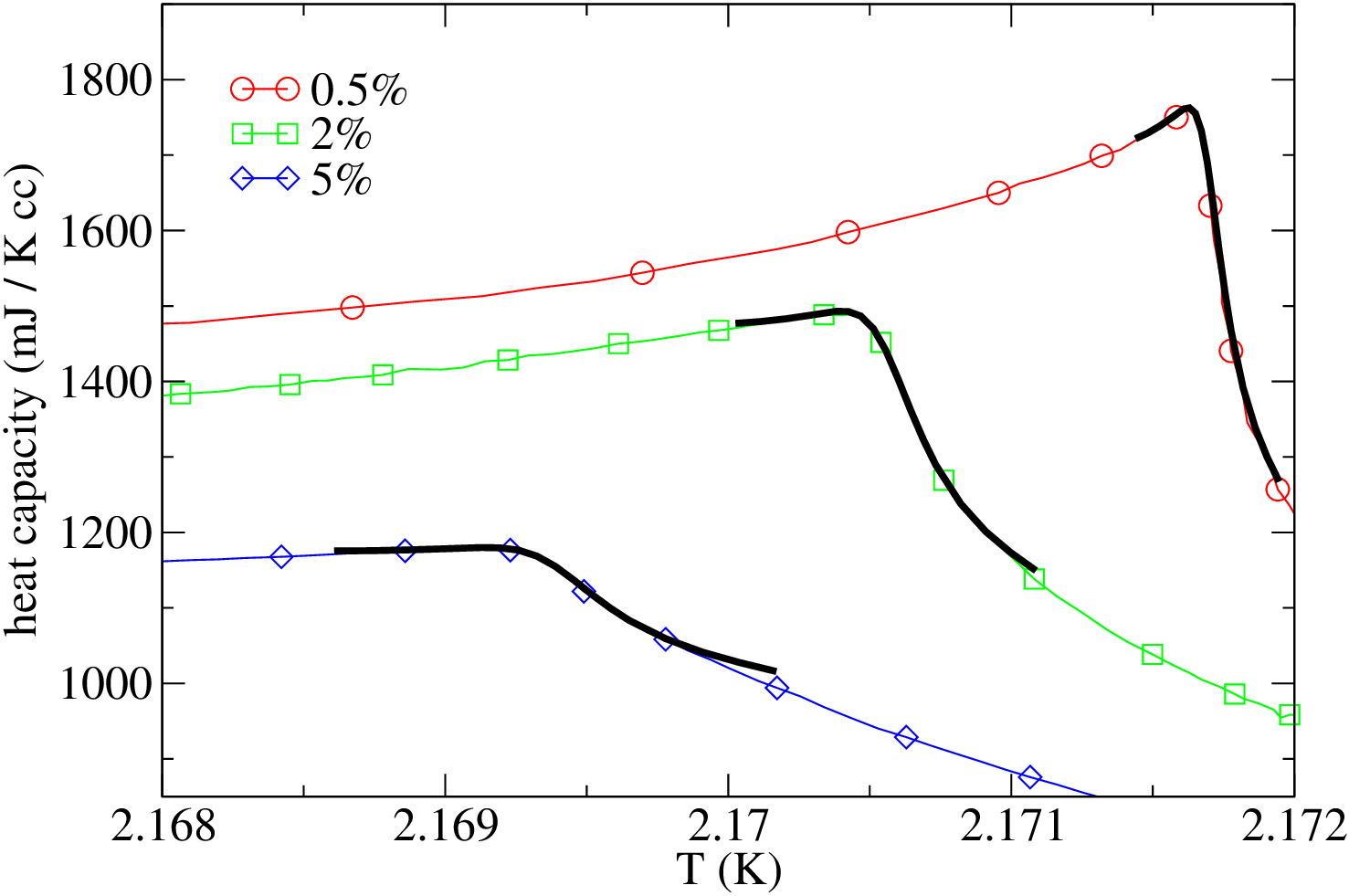}
\caption[]{(color online) Experimental data of the heat capacity for
three aerogel volume fractions, 0.5\%, 2\%, and 5\%, reproduced from
Ref.\ \cite{yoon} (curves with symbols), together with fits to the
scaling function (thick curves) given by Eq.\ (\ref{cfit}).  }
\label{fig:experiment}
\end{figure}

Further, we argue that the temperature dependence of the superfluid
condensate amplitude may contribute to the observed asymmetry of the
experimental heat capacity curves around $T_c$.  The fit presented in
Ref.\ \cite{yoon} to extract $\alpha$ from experiments assumes that
only the singular $c \sim |T-T_c|^{-\alpha}$ dependence on $T$ is
observed.  We propose that this might be too restrictive in cases when
$\alpha$ is negative, since the singular temperature dependence does
not clearly dominate the temperature variation as when $\alpha \gtrsim
0$.  We include the regular temperature dependence of the amplitude by
a linear approximation of the form $A=A_0+A_1T$, where $A_0,A_1$ are
constants \cite{ahlers84}.  This gives
\begin{equation}
c(T)= (A_0+A_1T) \left\{ 
L^{\alpha/\nu} \tilde{c}[L^{1/\nu}(T-T_c)]+b \right\},
\label{cfit}
\end{equation}
where the scaling function $\tilde{c}$ is determined in Fig.\
\ref{fig:Ycscaling}.  To fit Eq.\ (\ref{cfit}) to experiments we
assume $\alpha=-0.20, \nu=0.73$ and use $L,b,A_0,A_1$ as fit
parameters.  The results of the fits are shown in Fig.\
\ref{fig:experiment}.  The fit parameters are: $m=5\%: L=835, b= 5.03,
A_0=13592, A_1=-6150; m=2\%: L=1107, b=2.84, A_0=65914, A_1=-30098;
m=0.5\%: L=1900, b=1.98, A_0=193062, A_1=-88428$.  We thus obtain
quite good agreement in a narrow interval around the transition
between the experiment of Ref.\ \cite{yoon} and a fit to the simulated
scaling function, which fulfills hyperscaling.  The 5\% data curve
shows a slight deviation between theory and experiment on the high
temperature side of the transition in Fig.\ \ref{fig:experiment}.
This deviation occurs away from the transition temperature, where
additional temperature dependencies are expected.  Note that the
cutoff length depends roughly as $L \sim 1/ m^{1/3}$, which is
compatible with the interpretation as a crossover related to the
typical pore diameter in the aerogel.

{\em Discussion. --} From our finite size scaling analysis of MC data
for the 3D XY model in aerogel modeled as DLCA clusters, we obtain a
correlation length exponent $\nu\approx 0.73$.  This supports the
previous conclusion \cite{moon,vasquez} that the universality class of
the phase transition in the presence of DLCA clusters deviates from
the pure 3D XY universality class, where $\nu\approx 0.671$
\cite{campostrini}, even though the possibility of a slow crossover
towards the pure 3D XY universality class cannot entirely be ruled out.
In contrast to previous results in the literature \cite{yoon,vasquez},
our results suggest that the critical exponents are independent of
porosity and obey the Josephson hyperscaling relation $\alpha=2-d\nu$,
within the statistical uncertainty of the MC data.  We argue that the
formula $c= A_\pm|T-T_c|^{-\alpha_\pm}+B$, used to analyze experiments
in Ref.\ \cite{yoon} is not directly applicable in this case.  We
propose an alternative analysis procedure based on our simulation
result to fit the experimental heat capacity data at the transition.
This fit assumes that the experiment contains signatures of a finite
crossover length scale, presumably related to the typical aerogel pore
size, and of a temperature dependent superfluid condensate amplitude.

We thank Steve Girvin and Steve Teitel for helpful discussions.  This
work was supported by the Swedish Research Council, the G{\"o}ran
Gustafsson foundation, and the Swedish National Infrastructure for
Computing (SNIC).

\end{document}